\documentclass[useAMS,usenatbib]{mn2e}
\usepackage {graphicx}
\usepackage{natbib}
\usepackage{aas_journals}
\usepackage{color}
\usepackage{amsmath}
\newcommand{\mpc}{\rm {h^{-1}Mpc }}
\newcommand{\ltsima}{$\; \buildrel < \over \sim \;$}
\newcommand{\lsim}{\lower.5ex\hbox{\ltsima}}
\newcommand{\gtsima}{$\; \buildrel > \over \sim \;$}
\newcommand{\gsim}{\lower.5ex\hbox{\gtsima}}

\def\gtrsim{\mathrel{\hbox{\rlap{\hbox{\lower4pt\hbox{$\sim$}}}\hbox{$>$}}}}
\let\ga=\gtrsim
\def\lesssim{\mathrel{\hbox{\rlap{\hbox{\lower4pt\hbox{$\sim$}}}\hbox{$<$}}}}
\let\la=\lesssim

\newcommand{\LCDM}{$\Lambda$CDM}

\title[PS1-WISE-2MASS void]{Detection of a Supervoid Aligned with the Cold Spot of the Cosmic Microwave Background}
\author[Istv\'an Szapudi, Andr\'as Kov\'acs, Benjamin R. Granett, Zsolt Frei, Joe Silk, and the PS1 collaboration]{Istv\'an Szapudi$^{1}$, Andr\'as Kov\'acs$^{2,3,4}$, Benjamin R. Granett$^{5}$, Zsolt Frei$^{2,3}$, Joseph Silk$^6$, \newauthor
Will Burgett$^{1}$, Shaun Cole$^{7}$, Peter W. Draper$^{7}$, Daniel J. Farrow$^{7}$, Nicholas Kaiser$^{1}$,\newauthor 
Eugene A. Magnier$^{1}$, Nigel Metcalfe$^{7}$, Jeffrey S. Morgan$^{1}$, Paul Price$^{8}$, John Tonry$^{1}$,\newauthor Richard Wainscoat$^{1}$\\
$^{1}$ Institute for Astronomy, University of Hawaii 2680 Woodlawn Drive, Honolulu, HI, 96822, USA\\
$^{2}$ Institute of Physics, E\"otv\"os Lor\'and University, 1117 P\'azm\'any P\'eter s\'et\'any 1/A Budapest, Hungary\\
$^{3}$ MTA-ELTE EIRSA "Lend\"ulet" Astrophysics Research Group, 1117 P\'azm\'any P\'eter s\'et\'any 1/A Budapest, Hungary\\
$^{4}$ Institut de F\'isica d'Altes Energies, Universitat Aut\'onoma de Barcelona, E-08193 Bellaterra (Barcelona), Spain\\
$^{5}$ INAF OA Brera, Via E. Bianchi 46, Merate, Italy\\
$^{6}$ Department of Physics and Astronomy, The Johns Hopkins University, Baltimore MD 21218, USA\\
$^{7}$ Department of Physics, Durham University, South Road, Durham DH1 3LE, UK\\
$^{8}$ Department of Astrophysical Sciences, Princeton University, Princeton, NJ 08544}

\begin{document}

\date{Submitted 2014}

\pagerange{\pageref{firstpage}--\pageref{lastpage}} \pubyear{2012}

\maketitle

\label{firstpage}
\begin{abstract}
We use the WISE-2MASS infrared galaxy catalog matched with Pan-STARRS1 (PS1) galaxies to search for a supervoid in the direction of the Cosmic Microwave Background Cold Spot. Our imaging catalog has median redshift $z\simeq 0.14$, and we obtain photometric redshifts from PS1 optical colours to create a tomographic map of the galaxy distribution. The radial profile centred on the Cold Spot shows a large low density region, extending over 10's of degrees. Motivated by previous Cosmic Microwave Background results, we test for underdensities within two angular radii, $5^\circ$, and $15^\circ$.  The counts in photometric redshift bins show significantly low densities at high  detection significance, $\gtrsim 5 \sigma$ and $\gtrsim 6 \sigma$, respectively,  for the two fiducial radii. The line-of-sight position of the deepest region of the void is $z\simeq 0.15-0.25$. Our data, combined with an earlier measurement by \cite{GranettEtal2010}, are consistent with  a large $R_{\rm void}=(220 \pm 50)\mpc $ supervoid with $\delta_{m} \simeq -0.14 \pm 0.04$ centered at $z=0.22\pm0.03$. Such a supervoid, constituting at least a $\simeq 3.3\sigma$ fluctuation in a Gaussian distribution of the $\Lambda CDM$ model, is a plausible cause for the Cold Spot.
\end{abstract}
\begin{keywords}
surveys -- cosmology: observations -- large-scale structure of Universe -- cosmic background radiation
\end{keywords}

\section{Introduction}
\begin{figure*}
\begin{center}
\includegraphics[width=85mm]{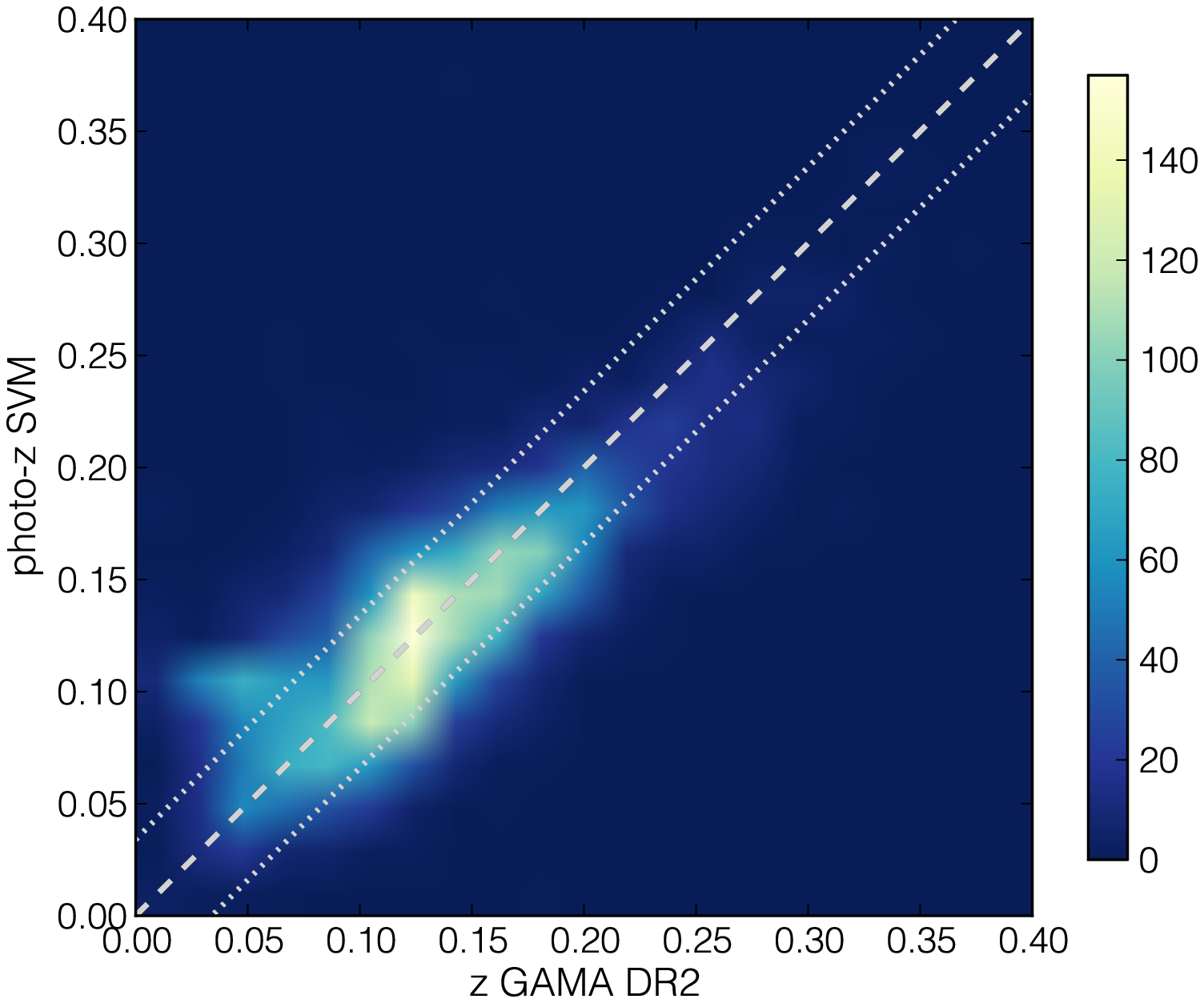}
\includegraphics[width=85mm]{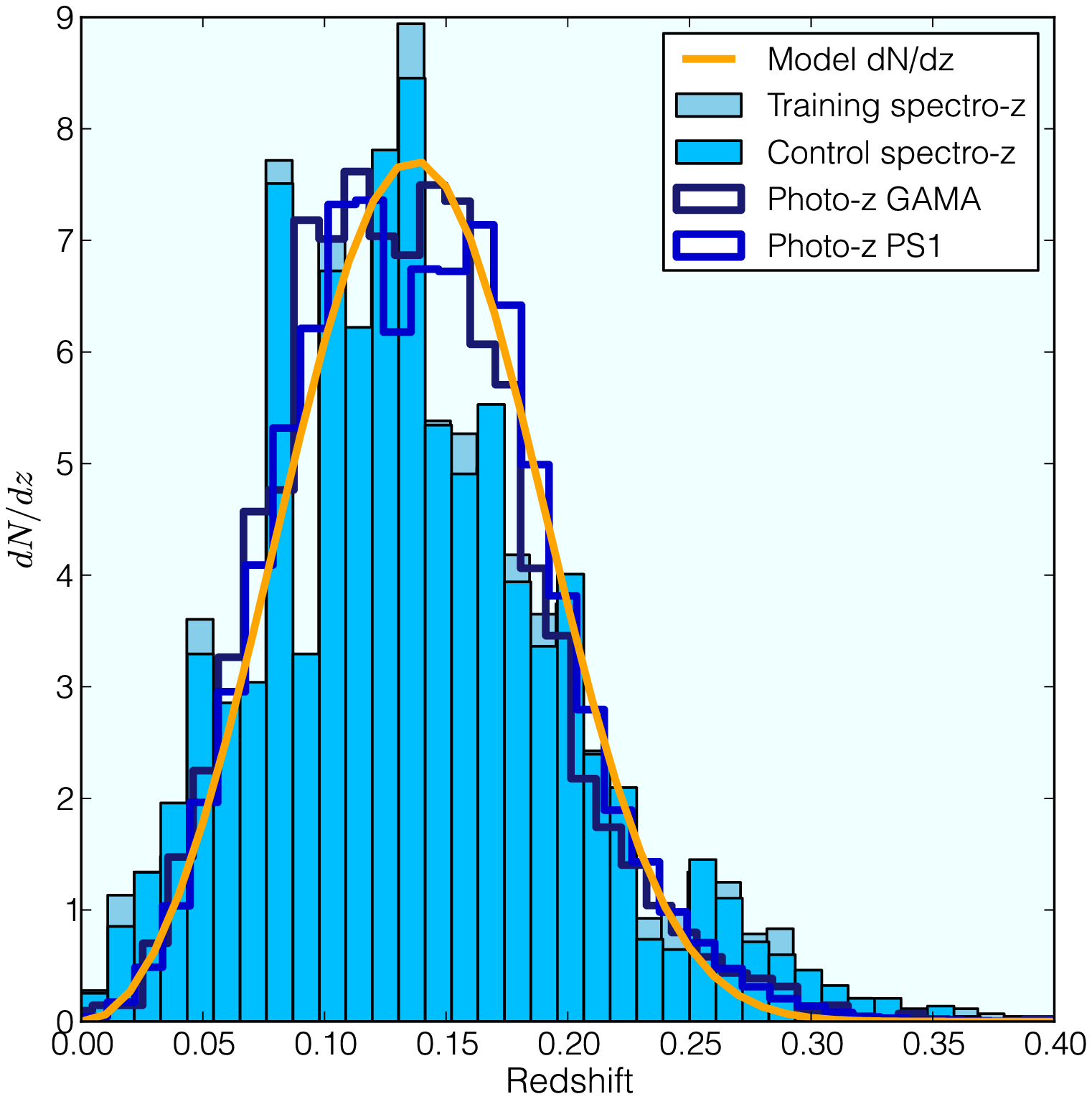}
\caption{The left panel shows the photo-z accuracy achieved by the SVM. Dotted lines indicate the $\sigma_{z}\approx 0.034$ $1\sigma$ error bars around the expectation. The right panel illustrates the normalized redshift distributions of our subsamples used in the photo-$z$ pipeline; training and control sets selected in GAMA, photo-$z$ distributions estimated for the WISE-2MASS-PS1-GAMA control sample, and photo-$z$'s of interest in the WISE-2MASS-PS1 matched area. The median redshift of all subsamples is $z\simeq 0.14$.}
\label{fig_pz}
\end{center}
\end{figure*}

The Cold Spot (CS) of the Cosmic Microwave Background (CMB) is an exceptionally cold $-70\mu$K area centred on $(l,b) \simeq (209^\circ,-57^\circ)$ Galactic coordinates. It was first detected in the Wilkinson Microwave Anisotropy Probe \citep{bennett2012} maps at $\simeq 3\sigma$ significance using wavelet filtering \citep{VielvaEtal2003, CruzEtal2004}. The CS is perhaps the most significant among  the ``anomalies'', potential departures from isotropic and/or Gaussian statistics, and all confirmed by {\it Planck} \citep{Planck23}. Explanations of the CS range from statistical fluke through hitherto undiscovered physics, e.g., textures \citep{CruzEtal2008,Vielva2010},  to the linear and nonlinear ISW effect \citep{SachsWolfe,ReesSciama} from a $\gtrsim 200\mpc$ supervoid centred on the CS  \citep{InoueSilk2006,InoueSilk2007,InoueEtal2010}. The latter would be readily detectable in large scale structure surveys thus motivating several observational studies.

A low density region approximately aligned with the CS was detected in a catalog of radio galaxies \citep{RudnickEtal2007}, although its significance has been disputed \citep{SmithHuterer2010}. A targeted redshift survey in the area \citep{BremerEtal2010} found no evidence for a void in the redshift range of $0.35 < z < 1$, while an imaging survey with photometric redshifts \citep{GranettEtal2010} excluded the presence of a large underdensity of $\delta\simeq -0.3$ between redshifts of $0.5 < z < 0.9$ and finding none at $0.3 < z < 0.5$. Both of these surveys ran out of volume at low redshifts due to their small survey area, although the data are consistent with the presence of a void at $ z < 0.3$ with low significance \citep{GranettEtal2010}. In  a shallow photometric redshift catalogue constructed from the Two Micron All Sky Survey \citep[2MASS]{2MASS} and SuperCOSMOS \citep{supercosmos} with a median redshift of $z=0.08$ an under-density was found \citep{francis2010} that can account for a CMB decrement of $\Delta T \simeq -7 \mu$K in the standard $\Lambda$-Cold Dark Matter (\LCDM) cosmology. While so far no void was found that could fully explain the CS, there is strong, $\gtrsim 4.4\sigma$, statistical evidence that superstructures imprint on the CMB as cold and hot spots \citep{GranettEtal2008,GranettEtal2009,PapaiSzapudi2010,Planck23,CaiEtal2013}. Note that the imprinted temperature in all of these studies is significantly colder than simple estimates from linear ISW \citep[e.g.,][]{RudnickEtal2007,PapaiSzapudi2010, PapaiEtal2011} would suggest.

The Wide-field Infrared Survey Explorer \citep[WISE]{WISE} all-sky survey effectively probes low redshift $z \leq 0.3$ unconstrained by previous studies. Using the WISE-2MASS all sky galaxy map of \cite{KovacsSzapudi2014} as a base catalog, we match a 1,300 $\sq^\circ$ area with the PV1.2 reprocessing of Pan-STARRS1 \citep[herafter PS1,][]{ps1ref}, adding optical colours for each objects. In the resulting catalog with photometric redshifts we test for the presence of a large low density region, a supervoid, centered on the CS. 
We defined the centre of the CS from the latest {\it Planck} results \citep{Planck24}. Based on the literature, we decided in advance to test for an underdensity at $5^\circ$ \citep{VielvaEtal2003, CruzEtal2004, RudnickEtal2007,GranettEtal2010,BremerEtal2010} and $15^\circ$ \citep{ZhangHuterer2010,InoueEtal2010} of radii. The fact that these values gleaned from CMB independently of our (large scale structure) data simplifies the interpretation of our results in the Bayesian framework, in particular, minimize any posteriori bias.

The paper is organised as follows. Data sets and map making algorithms are described in Section 2; our observational results are presented in Section 3; the final section contains a summary, discussion and interpretation of our results.

\section{Datasets and methodology}
Initially, we select galaxies from the WISE-2MASS catalog \citep{KovacsSzapudi2014}
containing sources to flux limits of $W1_{\rm WISE}$ $\leq $ 15.2 mag and $W1_{\rm WISE} - J_{\rm 2MASS}$ $\leq $ -1.7. We add a further limit of $J_{\rm 2MASS}$  $\leq $ 16.5 mag to ensure spatial homogeneity based on our experiments. This refinement shifts the median redshift of the sample to $z\simeq 0.14$. The catalog covers 21,200 $\sq^\circ$ after masking. We mask pixels with $E(B-V) \geq 0.1$, and regions at galactic latitudes $|b| < 20^{\circ}$ to exclude potentially contaminated regions near the Galactic plane \citep{SchlegelEtal1998}. These conservative limits result in a dataset deeper than the  2MASS Extended Source Catalog \citep[XSC]{jarrett2000} and more uniform than WISE \citep{KovacsEtAl2013}.
These galaxies have been matched with Pan-STARRS1 objects within a $50^\circ\times50^\circ$ area centred on the CS, except for a $\rm{Dec}\geq-28.0$ cut to conform to the PS1 boundary. We used PV1.2 reprocessing of PS1 in an area of 1,300 $\sq^\circ$.

\begin{figure*}
\begin{center}
\includegraphics[width=85mm]{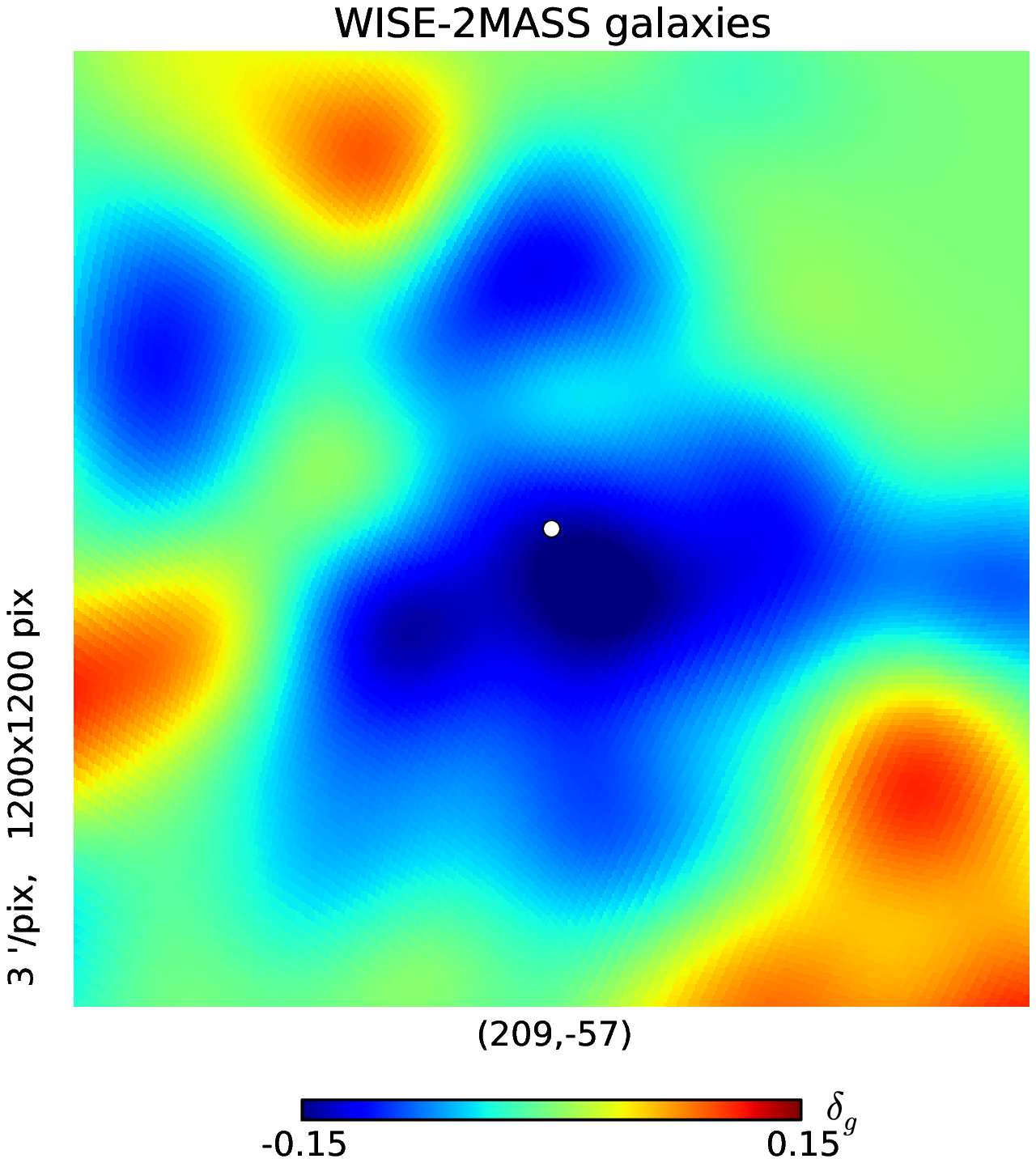}
\includegraphics[width=85mm]{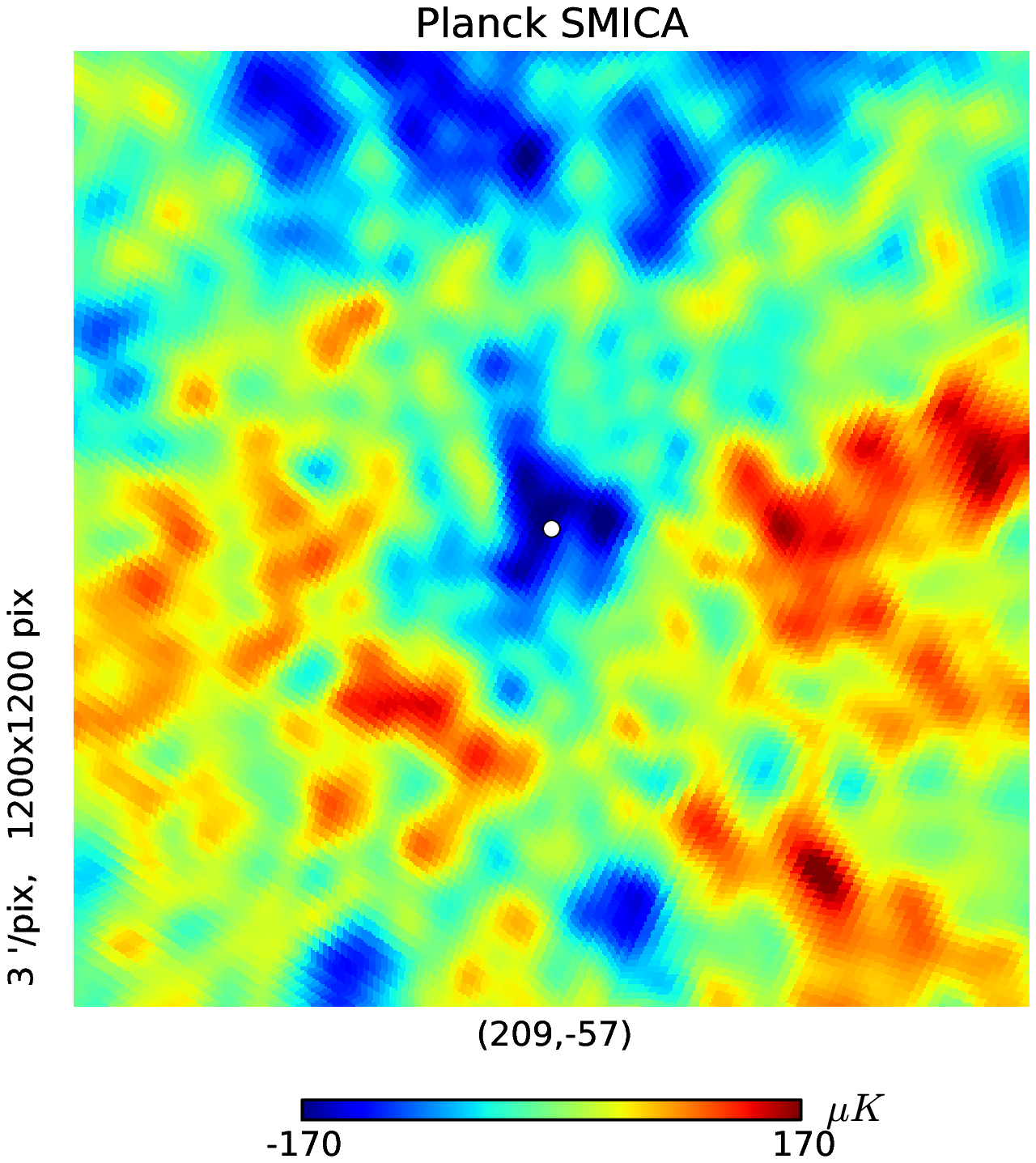}
\caption{ Gnomonic projections of the WISE-2MASS projected density map {\it (left)} and the Planck SMICA CMB map {\it (right)}. Both maps were created at $N_{side}=128$ resolution. We applied a Gaussian smoothing of $10^{\circ}$ ($2^{\circ}$) to the WISE-2MASS (Planck SMICA) map. White points indicate the center of the image, that is the center of the Cold Spot as defined in Planck 2013 results. XXIII.}
\label{fig_w2ms}
\end{center}
\end{figure*}

For matching we applied a nearest neighbor search using the {\tt scipy} kd-Tree algorithm with $1"$ matching radius, finding a PS1 pair for $86\%$ of the infrared galaxies, and resulting 73,100 objects in the final catalog. Galaxies without a PS1 match are faint in the optical, and predominantly massive early-type galaxies at $z > 1$ \citep{yanetal2012}.
For PS1, we required a proper measurement of Kron \citep{Kron1980} and PSF magnitudes in $g_{\rm P1}$, $r_{\rm P1}$ and $i_{\rm P1}$ bands that  were used to construct photometric redshifts (photo-$z$'s) with a Support Vector Machine (SVM) algorithm, and the {\tt python} {\tt Scikit-learn} \citep{scikit-learn} routines in regression mode.
The training and control sets were created matching WISE-2MASS, PS1, and the Galaxy and Mass Assembly \citep[GAMA,][]{gama} redshift survey. 
We chose a Gaussian kernel for our SVM and trained on $80\%$ of the GAMA redshifts, while the
rest were used for a control set. We empirically tuned the standard SVM parameters, finding the best performance when using $C=10.0$, and $\gamma = 0.1$. We characterize our photo-$z$ quality with the error $\sigma_{z} = \sqrt{\langle (z_{phot}-z_{spec})^{2} \rangle}$, finding $\sigma_{z}\approx 0.034$, as summarized in Figure \ref{fig_pz}.

\section{Results}

 The projected WISE-2MASS galaxy density field along with the {\it Planck} SMICA CMB map are shown in Figure \ref{fig_w2ms}. We have found that the most prominent large-scale underdensity found in WISE-2MASS is well aligned with the Cold Spot, although their sizes are different. Next we examine the radial statistics of this projected galaxy field, and apply photometric redshift techniques for a tomographic imaging of the region of interest. To avoid confusion, we will use the word ``significance'' to denote the significance of the detection of an underdensity, while ``rarity'' will denote the probability (expressed in $\sigma$'s) that the particular underdensity would appear in a cosmological random field.

\subsection{Significance and Rarity: 2D}
\label{ss_radial}
\begin{figure}
\begin{center}
\includegraphics[width=90mm]{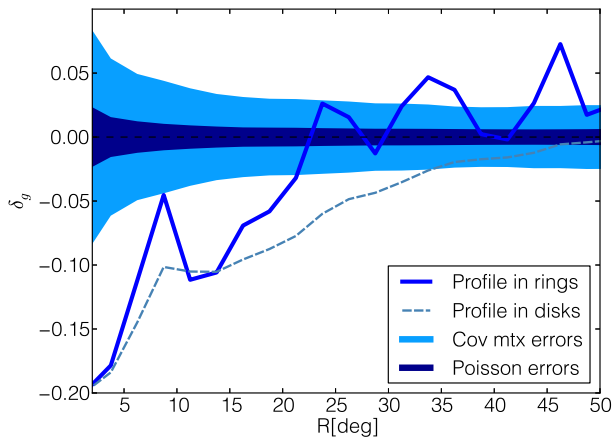}
\caption{ Radial galaxy density profile of the WISE-2MASS galaxy catalog, centred on the CS. The under-density is detected out to 10's of degrees in radius, consistent with an $r \approx 200\mpc$ supervoid with $\delta_{g} \simeq -0.2$ deepening in its centre. Note that the deeper central region is surrounded by a denser shell. }
\label{fig_prof}
\end{center}
\end{figure}

We first  study the Cold Spot region in projection, using the WISE-2MASS galaxies only. We measure radial galaxy density profiles in rings and disks centered on the CS in a bin size of $2.5^{\circ}$, allowing identification of relatively small-scale structures. In Figure \ref{fig_prof}, the dark shaded region represents Poisson fluctuations in our measurement in rings, calculated from the total number of galaxies in a ring. Figure \ref{fig_prof} shows a significant depression of sources. 
The size of the underdensity is surprisingly large: it is detected up to $\sim 20^\circ$ with high ($\gtrsim 5\sigma$)  detection significance. In addition, the profile has ring-like over density surrounding the CS region at large angular radii. This is consistent with a supervoid surrounded by a gentle compensation that converges to the average galaxy density at $\sim 50^\circ$ \citep[see e.g., ][]{PapaiEtal2011}.
At our pre-determined radii, $5^\circ$ and $15^\circ$, we have a signal-to-noise ratio $S/N \sim 12$ for detecting the rings.

\begin{figure*}
\begin{center}
\includegraphics[width=150mm]{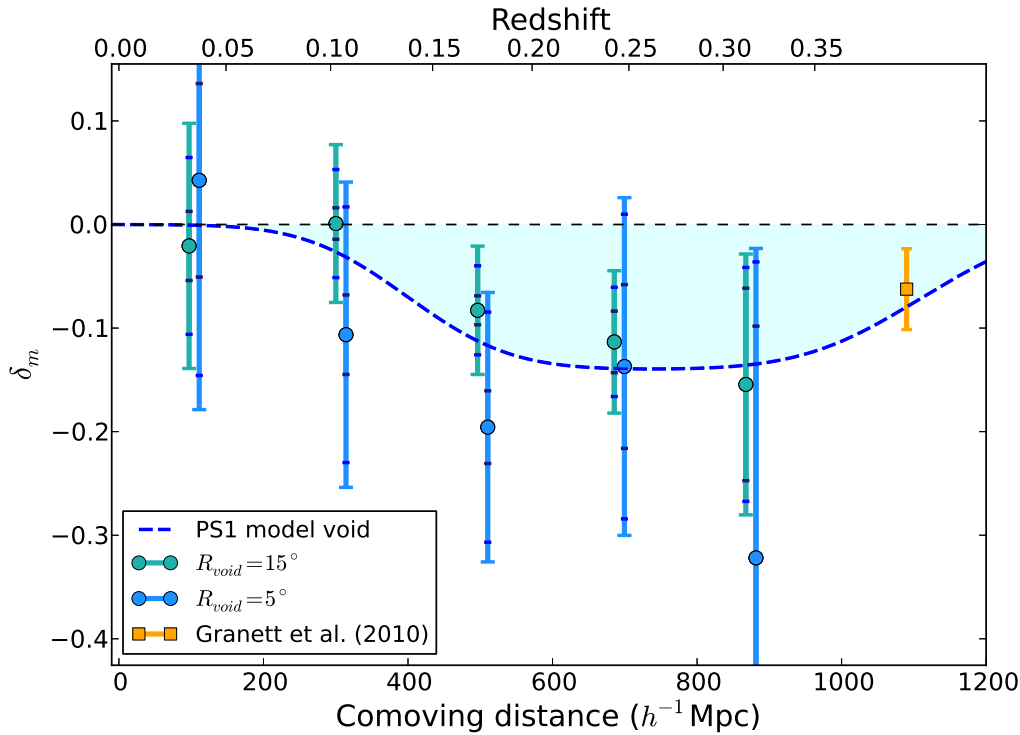}
\caption{Our measurements of the matter density in the line-of-sight using the $\Delta z = 0.07$ photo-$z$ bins we defined. We detected a significant depression in $\delta_{m}$ in $r=5^{\circ}$ and $15^{\circ}$ test circles. We used our simple modeling tool to examine the effects of photo-$z$ errors, and test the consistency of simple top-hat voids with our measurements. A data point by Granett et al (2010) accounts for the higher redshift part of the measurement. Dark blue (blue) stripes in error bars mark the contribution of Poisson (cosmic variance) fluctuations to the total error, while the additional part of the bars indicates the systematic effect of small survey coverage. See text for details.}
\label{fig_voids}
\end{center}
\end{figure*}

These results represent the detection significance of the underdensity calculated from Poisson errors. To quantify the cosmic rarity of the structure, we estimated the error bars arise from cosmic variance as well. Poisson fluctuations and cosmic variance errors are compared in Figure \ref{fig_prof}, corresponding to dark and light shaded regions. We created 10,000 Gaussian simulations of the projected galaxy map using the {\tt Healpix synfast} routine. As an input, we used a theoretical angular power spectrum assuming flat \LCDM~ cosmological model, and the redshift distribution of the WISE-2MASS sources. With the full covariance information, we evaluated a $\chi^{2}$ statistic for our radial density profile measurement compared to zero value in each bin. We have found $\chi^{2}=43.94$ for 24 degrees-of-freedom (the number of radial bins), i.e. $p=0.007$ or $\sim 3 \sigma$ characterizing the cosmic rarity of the supervoid in the (projected) concordance \LCDM\ framework.

\subsection{Significance: 3D}

We use the WISE-2MASS-PS1 galaxy catalog with photo-$z$ information to constrain the position, size, and depth of supervoids.
We count galaxies as a function of redshift in disks centred on  the CS at our fiducial angular radii,
$r=5^\circ$, and $15^\circ$, and compare the results to the average redshift distribution of our sample. Since the latter disks are cut by the PS1 mask, we always use the available area, and compensate accordingly. We fit the observed redshift distribution with a model $dN/dz \propto e^{-(z/z_{0})^{\alpha}} z^{\beta}$, estimating the parameters as $z_{0}=0.16$, $\alpha=3.1$, and $\beta=1.9$.
The average redshift distribution was obtained by counting all galaxies within our catalog outside the $15^{\circ}$ test circle, i.e. using 750 $\sq^\circ$, and errors of this measurement are propagated to our determination of the underdensity as follows. Our photo-$z$ bins were of width $\Delta z = 0.07$, and we compared the galaxy counts inside the test circles to those of the control area. We added an 
extra bin from the measurement of \cite{GranettEtal2010}  centered at $z=0.4$ in order to extend our analysis to higher redshifts in Fig. \ref{fig_voids}. 

Assuming accurate knowledge of the average density, the detection significance has Poisson statistics. However, given the fact that our photo-$z$ catalog is less than a factor of 2 larger than the area enclosed within the $15^\circ$ radius, we include error corresponding to the uncertainty of the average density due to Poisson and cosmic variance, as well as systematic errors. We estimate each term using simulations and the data.

\begin{figure}
\begin{center}
\includegraphics[width=85mm]{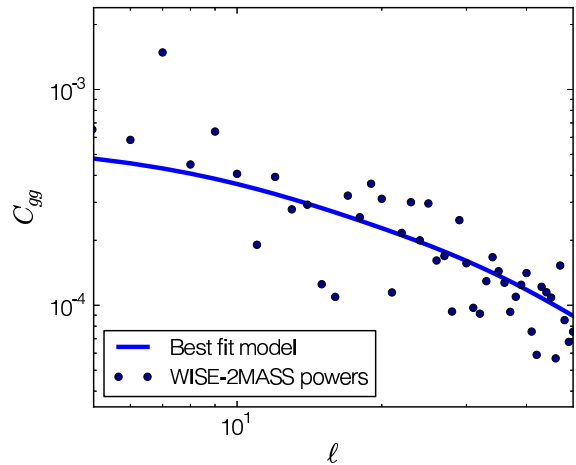}
\caption{Measurement of the angular power spectrum of the WISE-2MASS galaxies is presented along with the best fit theoretical model from concordance \LCDM~cosmology, and a best-fit model with bias $b_{g}=1.41\pm 0.07$. See text for details.}
\label{fig_bias}
\end{center}
\end{figure}

In order to create simulations of the density field, we first estimated the bias of the galaxy distribution. We modeled the angular power spectrum of the WISE-2MASS galaxy density map using the {\tt python} {\tt CosmoPy} package\footnote{{\texttt http://www.ifa.hawaii.edu/cosmopy/}}, and performed a measurement using {\tt SpICE} \citep{SzapudiB}. We assumed concordance flat \LCDM~ cosmological model with a fiducial value for the amplitude of fluctuations $\sigma_8=0.8$.  Then we carried out a $\chi^{2}$-based maximum likelihood parameter estimation, finding $b_{g}=1.41\pm 0.07$. The minimum value of $\chi_{min}^{2}=4.72$ is an excellent fit for $\nu=7$ degrees-of-freedom of our fitting procedure (8 bins in the angular power spectrum shown in Fig. \ref{fig_bias} and an amplitude parameter). 
This bias is comparable to earlier findings that measured the value of $b_{g}$ for 2MASS selected galaxies \citep{RassatEtal2003}, despite the additional uncertainty due to  that of $\sigma_8$. 
Using the bias, we generated $C_{gg}$ galaxy angular power spectra with {\tt CosmoPy} for the five photo-z bins applying a sharp cut to the full redshift distribution. As before, we assumed the concordance flat \LCDM~ cosmology. We then applied the same procedure as in Section \ref{ss_radial} using {\tt  synfast} for generating 1,000 random {\tt  HEALPix} simulations for each photo-$z$ bin.

The cosmic variance affecting the average density from using a small patch on the sky is characterized by estimating the variance of differences in mean densities estimated in the 
PS1 area and in full sky. In addition, we estimated a systematic errors by comparing cosmic variance from simulations to the variance of the average density of small patches measured in PS1 data. The extra variance corresponds to systematic errors, and possibly to any (presumably small) inaccuracy of our concordance cosmology and bias models.  The total error thus corresponds to the above three contributions shown in Figure \ref{fig_voids}. The procedure was repeated for each photo-$z$ bins.
We compared mean densities estimated in the part of the PS1 area used for obtaining the average density, and those measured in $R=5^{\circ}$ and $R=15^{\circ}$ circles. 

Qualitatively, simulations at lower redshifts contain stronger fluctuations on large scales, as the input power spectra contain higher powers for low-$\ell$. This effect is reflected in the systematic and cosmic variance error contributions we obtained, since the value of these corrections gradually decreases by $\sim 50 \%$ from bin 1 to bin 5. See Figure \ref{fig_voids} for details.

Using the above determined error bars, we find $S/N \sim 5$ and $S/N \sim 6$ for the deepest under-density bins for $r=5^{\circ}$ and $15^{\circ}$  characterizing our detection significance in 3D.

\subsection{Top Hat Supervoid Model and Rarity in 3D}
{To aid the interpretation of these results, we built toy models from top-hat voids in the $z$ direction and modeled the smearing by the photo-$z$ errors. The initial top hat  with three parameters, redshift ($z_{\rm void}$), radius ($R_{\rm void}$), and central depth ($\delta_{m}$), was smoothed using the distribution corresponding to the photometric redshift errors. The model redshift distribution was then multiplied with this smeared profile.

The void model can be compared to observations using a $\chi^{2}$-based maximum likelihood parameter estimation. We focus on the largest scale underdensity, therefore we only use the $r=15^{\circ}$ data, and replace our last bin with the measurement of \cite{GranettEtal2010}.  This gives $n=6$ bins with $k=3$ parameters, thus the degrees of freedom are $\nu = n-k=3$. 
We find a $\chi^{2}_{15^{\circ}} = 7.74$ for the null hypothesis of no void. The best fit parameters with marginalized errors are  $z_{\rm void}=0.22\pm 0.03$,  $R_{\rm void}=(220 \pm 50)\mpc $, and $\delta_{m}=-0.14\pm 0.04$ with $\chi^{2}_{min} = 3.55$.} Despite the simplicity of the toy model, the minimum chi-square indicates a good fit, expecting $\chi^{2}_{min} = \nu \pm \sqrt{2 \nu}$.
Nevertheless, more complexity is revealed by these counts, as bins 2-3 of the $r=5^{\circ}$ counts at redshifts $0.10\leq z \leq 0.15$ evidence the deepening of the supervoid in the center, or substructure.
For accurate prediction of the effect on the CMB, the density field around the CS region, including any substructure needs to be mapped precisely. This is left for future work, although we present a preliminary tomographic imaging of the region next. Nevertheless, using the above parameters and errors, we estimate that an underdensity is at least $3.3\sigma$ rare in a $\Lambda$CDM model with $\sigma_8 \simeq 0.8$, integrating the power spectrum to obtain the variance at $220\mpc$ and using Gaussian statistics for the probability. To get a lower bound on the rarity of the void, we used the fit parameters within their  $1-\sigma$ range always in the sense to \emph{increase} the likelihood of the underdensity in $\Lambda$CDM; thus the void we detected appears to be fairly rare.
Nevertheless, the top hat is an over-simplified toy model, thus the estimates based on it should be taken only as an initial attempt to interpret the detected supervoid in the concordance model framework.

\subsection{Tomographic Imaging}

\begin{figure}
\begin{center}
\includegraphics[width=85mm]{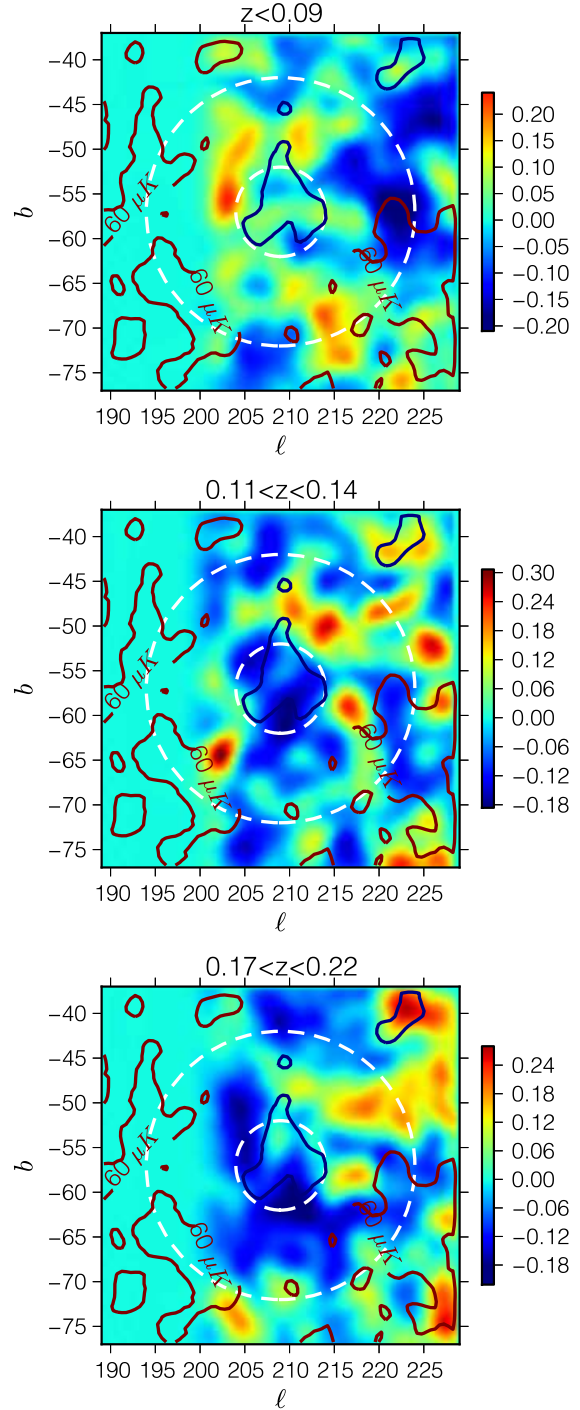}
\caption{Tomographic view of the CS region in $\delta_{m}$. The top panel appears to show a foreground over-density at the low redshift. A void is apparent at $0.11<z<0.14$ mostly inside the $5^{\circ}$ central region of the CS. The large under-density on the bottom panel at moderately higher redshifts may be slightly off centre with respect to the CS.}
\label{fig_visu}
\end{center}
\end{figure}

For three-dimensional impression of the galaxy distribution around the CS, we created maps in three photo-$z$ slices with a width of $z < 0.09$, $0.11 < z < 0.14$,  and $0.17 < z < 0.22$, and smoothed with a Gaussian at $2^{\circ}$ scales.  Then we over-plot the {\it Planck} SMICA CMB map as contours in Figure \ref{fig_visu}. The deepest part of the void appears to be close to the centre of the CS in the middle slice.

While photo-$z$ errors do not allow a fine-grained interpretation of the results, we observe a complex structure of voids, possibly a deeper, smaller void nested in a larger, shallower supervoid, or the deepening of a supervoid profile towards the middle. The foreground over density apparent in the first picture, especially the ``filament'' on the left side running along the PS1 survey boundary further complicates the picture. It is likely to be foreground, since it is more significant in the shallowest slice, gradually fading out at higher $z$. These tomographic maps  show a compensated surrounding overdense shell around the supervoid at $r \gtrsim 15^{\circ}$, which plausibly would have fragmented into galaxy clusters visible in the projected slices  as several "hot spots" surrounding the CS region. Note that  \cite{GurzadyanEtal2014} uses K-map statistics to Planck to show that the CS has a morphological structure similar to a void.

\section*{Discussion \& Conclusions}
Using our WISE-2MASS-PS1 data set, we detected a low density region, or supervoid, centred on the CS region: at $5^\circ$ and $15^\circ$ radii our detection significances are $5\sigma$ and $6\sigma$, respectively.  We measured the galaxy density as a function of redshift at the two predetermined radii. The galaxy underdensity is centred at $z \simeq 0.22$ for $15^\circ$, and even deeper around $z \simeq 0.15$ for $5^\circ$. The counts are consistent with a supervoid of size $R_{\rm void} \simeq 220 \mpc$ and average density $\delta_{g} \simeq -0.2$. It is noteworthy that this result is comparable to the local $300\mpc$ size underdensity claimed by \cite{KeenanEtal2012} with $\delta_g \simeq -0.3$.

We estimated the true underdensity of the supervoid, by modelling the angular power spectrum of the WISE-2MASS galaxy density map, finding $b_{g}=1.41\pm 0.07$. 
The resulting underdensity in the dark matter field, therefore, is $\delta = \delta_{g} / b_{g} \simeq -0.14\pm 0.04$ assuming a linear bias relation. Given the uncertainties of our toy model, we estimated that the supervoid we detected corresponds to a rare, at least  $3.3\sigma$, fluctuation in \LCDM, although the one $\sigma$ range of our measurements is also consistent with a void very unlikely in concordance models.
This agrees very well with our estimate that the underdensity found in the projected WISE-2MASS is a $3\sigma$ fluctuation compared to simple Gaussian simulations. Let's denote the probability of finding a Cold Spot on the CMB with $p_{CS}$, the probability of finding a void in LSS with $p_{void}$, and finally the probability of them being in alignment by chance with $p_{match}$. Let $H_1$ be the hypothesis, that the two structures are random fluctuations, and their alignment is random, and $H_2$ the hypothesis that the void is a random fluctuation {\em causing} the Cold Spot. The ratio of probabilities is $p_{H_2}/p_{H_1} = 1/(p_{CS} p_{match})$. For instance, conservatively, if the alignment is at the $\simeq 2^\circ$ level and the rarity of the CS is only $\simeq 2\sigma$, the ratio still overwhelmingly favors $H_2$. 
Thus chance alignment of two rare objects is not plausible, and a causal relation between the CS and the supervoid is more likely by a factor of at least $\simeq 20,000$.

Using \cite{RudnickEtal2007}, we estimate that the linear ISW effect of this supervoid is of order $-20$ $\mu$K on the CMB. The effect might be a factor or few higher if the size of the void is larger, if the compensation is taken into account \citep{PapaiEtal2011}, and/or if non-linear and general relativistic effects are included \citep[e.g., ][]{InoueSilk2006,InoueSilk2007}. Most recently,
\cite{FinelliEtal2014} attempted to fit a non-linear LTB model \citep{GBH2008} based on the projected profile in the WISE-2MASS catalog, and finds an effect not much larger than our initial estimate.

Superstructures affect several cosmological observables, such as CMB power spectrum and two and three point correlation functions \citep{MasinaNotari2009B,MasinaNotari2010}, CMB lensing \citep{MasinaNotari2009,DasSpergel2009}, 21-cm lensing \citep{KovetzKamionkowski2013}, CMB polarization \citep{VielvaEtal2011}, or cosmic radio dipole \citep{Schwarz2014}, and even B-mode polarization \citep{Bicep2014}. Furthermore, the other CMB anomalies associated with large-angle correlations \citep{AoE2005,copi2006,CopiEtal2013} should be revisited in light of these findings.

Our results suggest the connection between the supervoid and the CS, but further studies addressing the rarity of the observed supervoid observationally would be needed to firmly establish it. This needs a larger photometric redshift data set that will reach beyond $50^\circ$ radius, such as
PS1 with the second reprocessing thus improved calibration, and Dark Energy Survey \citep{DES}. As a first step, we smoothed the projected WISE-2MASS map with a $25^\circ$ Gaussian finding only one void as significant as the one we discovered in the CS region. This second void, to be followed up in future research and located near the constellation Draco, is clearly visible in the shallow 2MASS maps of  \cite{RassatEtal2013,francis2010} as a large underdensity, and in the corresponding reconstructed ISW map of \cite{RassatStarck2013} as a cold imprint. Therefore the Draco supervoid is likely to be closer thus smaller in physical size.
More accurate photometric redshifts, possibly with novel methods such as that of \cite{MenardEtal2013},  will help us to constrain further the morphology and the size of the supervoids, and a deeper data set would constrain their extent redshift space. Any tension with \LCDM, e.g. in the possible rarity of the observed supervoids, could be addressed in models of modified gravity, ordinarily screened in clusters, but resulting in an enhanced growth rate of voids as well as an additional contribution to the ISW signal.

\section*{Acknowledgments}
IS acknowledges NASA grants NNX12AF83G and NNX10AD53G. AK and ZF acknowledge support from OTKA through grant no. 101666, and AK acknowledges support from the Campus Hungary fellowship programme, and the Severo Ochoa fellowship programme. BRG acknowledges support from the European Research Council Darklight ERC Advanced Research Grant (\# 291521). We used HEALPix \citep{healpix}. The Pan-STARRS1 Surveys (PS1) have been made possible through contributions by the Institute for Astronomy, the University of Hawaii, the Pan-STARRS Project Office, the Max-Planck Society and its participating institutes, the Max Planck Institute for Astronomy, Heidelberg and the Max Planck Institute for Extraterrestrial Physics, Garching, The Johns Hopkins University, Durham University, the University of Edinburgh, the Queen's University Belfast, the Harvard-Smithsonian Center for Astrophysics, the Las Cumbres Observatory Global Telescope Network Incorporated, the National Central University of Taiwan, the Space Telescope Science Institute, and the National Aeronautics and Space Administration under Grant No. NNX08AR22G issued through the Planetary Science Division of the NASA Science Mission Directorate, the National Science Foundation Grant No. AST-1238877, the University of Maryland, and the Eotvos Lorand University (ELTE).

\bibliographystyle{mn2e}
\bibliography{refs}
\end{document}